# Accurate myocardial T1 mapping at 5T using an improved MOLLI method: A validation study

Linqi Ge[a,b,c,1], Yinuo Zhao[a,b,1], Yubo Guo[e,1], Yuanyuan Liu[a,b], Yihang Zhou[b,d], Haifeng Wang[a,b], Dong Liang[b,d], Hairong Zheng[a,b], Yining Wang[e,*], Yanjie Zhu[a,b,**]

[a] Paul C. Lauterbur Research Center for Biomedical Imaging, Shenzhen Institutes of Advanced Technology, Chinese Academy of Sciences, Shenzhen, Guangdong, China

[b] State Key Laboratory of Biomedical Imaging Science and System, Shenzhen Institutes of Advanced Technology, Chinese Academy of Sciences, Shenzhen, Guangdong, China

[c] University of Chinese Academy of Sciences, Beijing, China

[d] Research Center for Medical AI, Shenzhen Institutes of Advanced Technology, Chinese Academy of Sciences, Shenzhen, Guangdong, China

[e] Department of Radiology, State Key Laboratory of Complex Severe and Rare Diseases, Peking Union Medical College Hospital, Chinese Academy of Medical Sciences and Peking Union Medical College, Beijing, China

[*] **Corresponding author.** Department of Radiology, Peking Union Medical College Hospital, No.1, Shuaifuyuan, Dongcheng District, Beijing 100730, China.

[**] **Corresponding author.** Paul C. Lauterbur Research Center for Biomedical Imaging, Shenzhen Institutes of Advanced Technology, Chinese Academy of Sciences, Shenzhen, China.

**E-mail addresses:** wangyining@pumch.cn (Y. Wang), yj.zhu@siat.ac.cn (Y. Zhu).

[1]**Co-first authors.**

**ABSTRACT**

**Background:** Accurate myocardial T1 mapping at 5T remains a technical challenge due to field inhomogeneity and prolonged T1 values. The aim of this study is to develop an accurate and clinically applicable myocardial T1 mapping technique for 5T magnetic resonance imaging (MRI) systems and validate its performance in a multicenter study.

**Methods:** The proposed method is based on a 5-(3)-3 Modified Look-Locker Inversion Recovery (MOLLI) sequence, dubbed combined-correction MOLLI (coMOLLI), which corrects for both inversion efficiency and readout-induced signal disturbances. Specifically, coMOLLI employs a radiofrequency spoiled gradient recalled echo (GRE) readout rather than the commonly used balanced steady state free precession (bSSFP) readout. Its signal evolution is modeled to estimate T1 values, which incorporates both inversion efficiency and readout disturbances to improve fitting accuracy. To further enhance accuracy, the inversion pulse was redesigned under hardware constraints and the observed $B_0$ and $B_1$ variations over the heart at 5T, using adiabatic hyperbolic secant (HSn) and tangent/hyperbolic tangent (Tan/Tanh) pulses. The method was validated in phantom experiments, as well as in 21 healthy volunteers and 9 patients.

**Results:** The optimized inversion pulse at 5T is the Tan/Tanh pulse with $A$ = 10 kHz, $K_s$ = 4, $k$ = 22, and $T_p$ = 8 ms. In phantom studies, coMOLLI showed high accuracy versus reference inversion recovery - fast spin echo (IR-FSE), yielding relative errors within 5% for all nine vials. In vivo studies, the average native myocardial T1 values across 21 healthy volunteers were 1468 ± 48 ms, 1514 ± 39 ms, and 1545 ± 50 ms, and blood T1 values were 2182 ± 132 ms, 2124 ± 153 ms, and 2131 ± 158 ms for apical, middle, and base slices, respectively.

**Conclusion:** The coMOLLI method demonstrated high accuracy in phantom studies and feasibility in vivo studies. By adopting the widely used 5-(3)-3 MOLLI acquisition scheme, it shows potential for clinical cardiac imaging at 5T.

**Abbreviations:**

| | |
|---|---|
| BPM | beats per minute |
| bSSFP | balanced steady state free precession |
| ECG | electrocardiogram |
| ECV | extracellular volume fraction |
| FA | flip angle |
| FOV | field-of-view |
| FSE | fast spin echo |
| Gd | gadolinium |
| GRAPPA | generalized autocalibrating partially parallel acquisition |
| GRE | radiofrequency spoiled gradient recalled echo |
| IR | inversion recovery |
| LGE | late gadolinium enhancement |
| LV | left ventricle/left ventricular |
| MATLAB | MathWorks, Inc., Natick, Massachusetts, USA |
| MINOCA | myocardial infarction with no obstructive coronary arteries |
| MOLLI | Modified Look-Locker Inversion recovery |
| MR | magnetic resonance |
| MRI | magnetic resonance imaging |
| mSASHA | multiparametric saturation recovery single-shot acquisition |
| NICM | non-ischemic cardiomyopathy |
| PDw | proton density weighted |
| RF | radiofrequency |
| ROI | region-of-interest |
| SAPPHIRE | saturation pulse prepared heart rate independent inversion recovery |
| SAR | specific absorption rate |
| SASHA | saturation recovery single-shot acquisition |
| SD | standard deviation |
| SNR | signal-to-noise ratio |
| T | Tesla |
| T1w | T1 weighted |
| TE | echo time |
| TI | inversion time |
| TR | repetition time |

## 1. Introduction

In the past few years, magnetic resonance (MR) myocardial T1 mapping has become an essential tool for the direct quantification of myocardial tissue characteristics [1]. Compared to late gadolinium enhancement (LGE) [2], myocardial T1 mapping, as well as its derived parameter, extracellular volume fraction (ECV) [3] demonstrates superiority in evaluating the degree of extracellular matrix expansion and delivers more accurate results in diffuse myocardial fibrosis detection [4]. In recent years, the development of high-field whole-body magnetic resonance imaging (MRI) systems exceeding 3T, i.e., 5T systems, has attracted increasing interest due to their potential signal-to-noise ratio (SNR) advantages. Several pilot studies have also demonstrated that 5T can provide higher image resolution and improved image quality in cardiac imaging [5-7]. However, accurate myocardial T1 mapping at 5T remains a technical challenge, requiring further development for robust clinical use.

Several techniques have been developed for myocardial T1 mapping at 3T and 1.5T MR systems. The most widely adopted technique is the Modified Look-Locker Inversion Recovery (MOLLI) [8]. It acquires a series of T1 weighted (T1w) images at different inversion times (TIs) following an inversion pulse, obtaining the pixel-wise T1 relaxation times by fitting the signal recovery curve to a three-parameter model. However, MOLLI requires data acquisition over multiple heartbeats, making the derived T1 values susceptible to heart rate variations. To address these issues, saturation recovery single-shot acquisition (SASHA) [9] replaces inversion recovery (IRs) with saturation pulses to eliminate heart rate dependence, yet its smaller dynamic signal range during recovery leads to a reduced SNR in the obtained images as well as the derived T1 maps. This drawback becomes even worse at 5T due to the longer T1 values. Saturation pulse prepared heart rate independent inversion recovery (SAPPHIRE) [10] improves accuracy for long T1 values by integrating inversion and saturation recovery, though its complexity limits clinical applications. The slice-interleaved T1 (STONE) sequence [11] provides accurate T1 mapping with high SNR, yet requires simultaneous acquisition of multiple slices. Model-based T1 mapping [12] uses continuous acquisition via radial trajectories after a single inversion pulse, and then reconstructs T1 maps directly from the acquired k-space data using a joint sparsity constrained model. This approach is resistant to motion artifacts, but its complex reconstruction process limits its clinical applicability.

Considering all trade-offs among the above techniques, we selected the MOLLI-based method as the baseline method. MOLLI remains the most widely adopted clinical technique for myocardial T1 mapping due to its robustness and reliability. However, its assumptions do not hold at 5T, where several key factors contribute to T1 inaccuracy, as outlined below. (1) The inversion process is not always perfect. (2) MOLLI assumes full recovery between pulses, which is hard to achieve in tissues with prolonged T1 at 5T or high heart rates. (3) MOLLI utilizes a fitting model initially developed for continuous radiofrequency spoiled gradient recalled echo (GRE) readouts [12], which leads to errors dependent on tissue T1 and T2 properties.

In this work, we aim to develop an accurate and clinically applicable myocardial T1 mapping technique for 5T systems, dubbed combined-correction MOLLI (coMOLLI), which corrects both inversion efficiency and signal disturbance caused by readout. Specifically, image acquisition is performed using a 5-(3)-3 MOLLI sequence with a GRE readout, while the inversion pulse is redesigned using adiabatic hyperbolic secant (HSn) and tangent/hyperbolic tangent (Tan/Tanh) pulses to improve inversion efficiency at 5T. Then, the T1 map is estimated using a scan-specific fitting model, similar to the Instantaneous Signal Loss Simulation (InSiL) [13] approach, which involves the effects of inversion efficiency, readout gradients, and the subject's heart rate. The proposed method was first validated via a phantom study, and then evaluated in vivo at two imaging centers, including 21 healthy volunteers and 9 patients with cardiac diseases. Results demonstrate that our technique achieves high accuracy in phantoms and is feasible for myocardial T1 mapping at 5T.

## 2. Methods

### 2.1 Sequence

The coMOLLI sequence was implemented on a 5T system (Jupiter, United Imaging Healthcare, China). The timing diagram of the sequence is shown in Fig. 1. It uses a 5-(3)-3 MOLLI scheme consisting of two consecutive electrocardiogram-gated (ECG-gated) inversion recovery modified Look-Locker blocks. A three-heartbeat rest period is inserted between the two inversion pulses to allow recovery of longitudinal magnetization before the second inversion. Data acquisition is performed using GRE readout in the mid-diastolic cardiac phases, synchronized via ECG-gating. We use GRE readout instead of the commonly used balanced steady state free precession (bSSFP), as bSSFP is prone to dark banding artifacts and has high specific absorption rate (SAR), making it less suitable for scanners with main magnetic field strengths above 3T (see Supplementary Fig. S1).

**2.2 T1 fitting**

To account for inversion imperfections, the signal evolution is modeled as follows:

(1) The magnetization after the inversion pulse is given by:

$$M_{inv}(k)^{+} = -\delta \cdot M_{inv}(k) \tag{1}$$

where $M_{inv}(k)$ and $M_{inv}(k)^{+}$ denote the longitudinal magnetization immediately before and after the $k$-th inversion pulse, respectively. $\delta$ is the inversion efficiency, and $k = 1, 2$ represents the inversion index. The initial signal $M_{inv}(1) = M_0$ represents the equilibrium longitudinal magnetization, hence $M_{inv}(1)^{+} = -\delta \cdot M_0$.

(2) We assume that disturbances caused by readouts are instantaneous and parameterized by a correction factor $C$ $(0 < C < 1)$:

$$M_{ro}(j)^{+} = (1 - C) \cdot M_{ro}(j) \tag{2}$$

where $j = 1,2,\ldots,5$ and $j = 6, 7, 8$ denote the readout events after the first and second inversion pulse, respectively, which is consistent with our deliberately designed 5-(3)-3 MOLLI sequence. The subscript "ro" indicates readout events. $M_{ro}(j)$ and $M_{ro}(j)^{+}$ represent the signal immediately before and after the $j$-th single-shot acquisition, respectively.

(3) Based on the above, the longitudinal magnetization evolution can be described through a series of relaxation equations. In particular, the first acquisition process is expressed as follows:

$$M_{ro}(1) = M_0 + [M_{inv}(1)^{+} - M_0]\, exp\left(-\frac{\Delta T_1}{T_1}\right) \tag{3}$$

$$M_{ro}(1)^{+} = (1 - C)M_0\left[1 - [\delta + 1]\, exp\left(-\frac{\Delta T_1}{T_1}\right)\right] \tag{4}$$

where $M_{ro}(1)$ and $M_{ro}(1)^{+}$ denote the signal immediately before and after the first acquisition, respectively. $\Delta T_1$ denotes the time interval between the first inversion pulse and the acquisition of the k-space center line in the first readout. The signal at subsequent acquisition steps can be computed recursively as follows:

$$M_{all}(i) = M_0 + [M_{all}(i-1)^{+} - M_0]\, exp\left(-\frac{\Delta T_i}{T_1}\right) \quad (5)$$

where $i = 1, 2, ..., 10$ denotes either the inversion event or the readout event, and $\Delta T_k$ denotes the time interval between an inversion pulse and the subsequent imaging acquisition, or between two consecutive imaging acquisitions.

(4) According to the above equations, the MR signals at all timepoints can be simulated, which contain four unknowns $M_0, T_1, \delta,$ and $C$. These unknowns can be estimated by minimizing the mean-squared errors between the simulated and measured signals [14]. To reduce computational complexity and improve fitting robustness, the inversion efficiency $\delta$ is pre-calculated using a pre-scan, as described below. Then, a three-parameter nonlinear optimization equation is employed to estimate the remaining unknowns $M_0, T_1,$ and $C$ for each pixel as follows:

$$[M_0, T_1, C] = argmin\left\{\sum_{j=1}^{J}\left(M_{ro}(j) - S(j)\right)^2\right\}, \qquad where\, J = 8 \quad (6)$$

where $S(j)$ denotes the actual measured signal, and $M_{ro}(j)$ denotes the simulated signal. In this study, the Levenberg-Marquardt algorithm [15] is used to solve the optimization problem of Eq.(6).

**2.3 Optimization of Inversion Pulse**

The inversion efficiency $\delta$ is essential for accurate myocardial T1 mapping [16]. Due to the SAR limitation at 5T, the maximum achievable $B_1$ value is 10.6 μT (about 450 Hz), which is much lower than the $B_1$ strength required for the optimal adiabatic inversion pulse for myocardial T1 mapping at 3T [16]. Therefore, the inversion pulse at 5T needs to be redesigned to optimize inversion efficiency.

Two adiabatic inversion pulse designs considered in this study were HSn [17-20] and Tan/Tanh [18,19,21]. For HSn design, the formula is given:

$$\begin{cases} \omega_1(t) = B_1\, sech\Big(\beta\big(2t/T_p - 1\big)^n\Big) \\ \Delta\omega(t) = A\, tanh\Big(\beta\big(2t/T_p - 1\big)^n\Big) \end{cases}, \qquad \big(0 < t \leq T_p\big) \tag{7}$$

where $T_p$ is the pulse duration, $A$ denotes the frequency sweep amplitude, and $\beta$ and n denote pulse shape parameters. The parameter ranges were: $n$ = 1, 2, 4, and 8, $A$ = 100−1500 Hz in 100 Hz steps, and $\beta = asech(x)$ with $x$ = 0.005−0.02 in steps of 0.005. For Tan/Tanh design, the formula is given by:

$$\begin{cases} \omega_1(t) = B_1\, tanh\big(2\xi t/T_p\big) \\ \Delta\omega(t) = A\Big(tan\Big(\kappa\big(2t/T_p - 1\big)\Big)/tan(\kappa)\Big) \end{cases}, \qquad \big(0 < t \leq T_p\big) \tag{8}$$

where $\xi$ and $\kappa$ denote pulse shape parameters. The parameter ranges of Tan/Tanh were: $A$ = 4000−15000 Hz in 500 Hz steps, tan ($\kappa$) = 8−30 in steps of 2, and $\xi$ = 2−20 in steps of 2. The ranges and steps of radiofrequency (RF) parameters were determined based on previous studies [16,22]. $B_1$ field strength was set to the maximum achievable value of 450 Hz for all designs. Considering the limits on SAR and achievable $B_1$, we empirically set the RF duration ($T_p$) range from 8 to 30 ms in 1 ms increments for both HSn and Tan/Tanh. To determine $B_1$ and $B_0$ variations at 5T, $B_1$ and $B_0$ field maps were acquired in the short-axis view using a small cohort of five volunteers on a 5T system. Based on the observed variations over the heart, the $B_0$ range was set from -250 Hz to 250 Hz for pulse optimization. The $B_1$ range was set as 50%−120% relative to the nominal value. As the used $B_1$ was 450 Hz, the corresponding $B_1$ range was 225−540 Hz for pulse optimization.

The inversion efficiency $\delta$ for a variety of inversion pulse designs was calculated using the Bloch simulation with T1 = 1500 ms and T2 = 40 ms. The simulation was conducted with the mri-rf package in the Michigan Image Reconstruction Toolbox (MIRT) [23]. The equilibrium magnetization $M_0$ was set to 1, and $\delta$ was calculated as the ratio of the simulated longitudinal magnetization after the inversion pulse to $M_0$. The $\delta$ across the $B_0$ and $B_1$ imperfection ranges were averaged, and the pulse parameter combination with the highest average $\delta$ was identified as the optimized set of parameters.

### 2.4 Inversion efficiency estimation

In previous studies, the “MOLLI + $M_0$” sequence was proposed to estimate the inversion efficiency $\delta$ by acquiring an additional proton density weighted (PDw) image after the MOLLI readouts [8]. However, at 5T, such inline acquisition introduces a longer scan duration and extended breath-hold time, especially in the presence of prolonged T1 values. Therefore, in this study, a separate sequence was employed to estimate $\delta$.

The timing diagram of this sequence is shown in Fig. 2. It first acquires a PDw image ($I_0$) using a single-shot GRE readout. After a five-heartbeat interval to allow full recovery of the magnetization, an inversion pulse is applied, immediately followed by a second acquisition ($I_{IR}$). Myocardial regions were manually delineated on both images, and the average signal within each region-of-interest (ROI) was calculated. Then $\delta$ was measured as the ratio of the average myocardial signals of $I_{IR}$ and $I_0$. The mean $\delta$ across 21 healthy volunteers was 0.855, which was subsequently used for all T1 fitting in Eq.(6).

### 2.5 Phantom study

All imaging experiments were performed on a 5T MR system (Jupiter, United Imaging Healthcare, China). The phantom comprises nine vials, which were made by $NiCl_2$-doped agarose gel with varying concentrations to mimic different cardiac compartments [24]. The T1 values of the phantom were first measured using the inversion recovery - fast spin echo (IR-FSE) sequence with a local transmit and 48-channel receiver head coil. The imaging parameters of the IR-FSE sequence were: repetition time (TR)/echo time (TE) = 15 s/9.56 ms, field-of-view (FOV) = 320 × 320 mm$^2$, acquisition matrix = 320 × 256, slice thickness = 5 mm, bandwidth = 260 Hz/pixel, and TIs = 75 ms, 100 ms, 125 ms, 150 ms, 200 ms, 300 ms, 500 ms, 800 ms, 1000 ms, 1500 ms, 2000 ms, 2500 ms. The T1 values were fitted with a three-parameter model from images obtained using the IR-FSE sequence and served as the gold standard.

The imaging parameters of coMOLLI were: TR/TE = 4.35 ms/1.59 ms, FOV = 150 × 150 mm$^2$, acquisition matrix = 128 × 128, slice thickness = 8 mm, flip angle (FA) = 7°, bandwidth = 800 Hz/pixel, and TIs = 155 ms and 235 ms. Image acceleration was performed using generalized autocalibrating partially parallel acquisition (GRAPPA) with an acceleration factor R = 2 and 24 calibration lines. The simulated heart rate was 75 beats per minute (BPM).

Accuracy was defined as the relative error (%) between the T1 values measured by coMOLLI and those obtained using IR-FSE [25]. Agreement between the two methods was further assessed using a Bland−Altman analysis. Precision was quantified per vial as the standard deviation (SD) of pixel-wise T1 values within the vial ROI [26,27].

**2.6 In vivo study**

The in vivo study was conducted in two centers: United Imaging Healthcare and Peking Union Medical College Hospital. The experiments were approved by the Institutional Review Boards (IRBs) of both centers, and informed consent was obtained from each participant before the scan. A total of 21 healthy volunteers (12 males and 9 females, aged 34 ± 26 years) and 9 patients with cardiac diseases (5 males and 4 females, aged 46 ± 31 years) were recruited.

Healthy volunteers were recruited from both centers: 7 from United Imaging Healthcare and 14 from Peking Union Medical College Hospital. Inclusion criteria for healthy subjects included no history or symptoms suggestive of cardiovascular disease and no contraindications to MRI. Patients were recruited exclusively from Peking Union Medical College Hospital. Inclusion criteria for patients included referral for clinical cardiac MRI due to suspected or confirmed cardiovascular disease. All patients were required to be able to comply with breath-holding instructions during image acquisition. The specific cardiomyopathies of these patients are summarized in Table 1.

For healthy volunteers, native T1 maps were acquired at three short-axis slices, namely apex, middle, and base. For patients, both native and post gadolinium (Gd) infusion T1 maps were acquired at the same three short-axis slices. Imaging parameters were: TR/TE = 3.99 ms/1.453 ms, FOV = 256 × 228 mm$^2$, acquisition matrix = 256 × 148, slice thickness = 8 mm, FA = 7°, bandwidth = 800 Hz/pixel, and TIs = 155 ms and 235 ms. The acquisition window of each image was 295 ms. A 24-channel phased-array coil was used for signal reception. All T1w images were first registered using a symmetric diffeomorphic registration algorithm [28] implemented on the scanner, and then fitted pixel-wise using Eq.(6) to obtain the T1 map.

For the in vivo analysis, the endocardial and epicardial contours of the left ventricle (LV) were manually delineated on the T1 maps to outline the myocardial ROI. The mean myocardial T1 value within each ROI was calculated from the T1 map. In addition, blood T1 is also measured. Sex-based comparisons of global T1 values among healthy volunteers were performed using two-sided Welch t-tests. Precision was defined as the SD of myocardial T1 values within each ROI [26,27], and was averaged across the 21 volunteers, yielding one precision value per slice level. All analyses were performed in MATLAB (R2023a; MathWorks, Inc., Natick, Massachusetts, USA).

## 3. Results

### 3.1 Simulations

For HSn, the best design is achieved with $\beta$ = 0.02, $A$ = 0.5 kHz, $power$ = 2, and $T_p$ = 10 ms, yielding a mean $\delta$ = 0.8916 across $\Delta B_0$ = ±250 Hz and $B_1$ variations from 50% to 120% relative to the nominal $B_1$ value (450 Hz). For Tan/Tanh, the best design is $A$ = 10 kHz, $K_s$ = 4, $k$ = 22, and $T_p$ = 8 ms yielding a mean $\delta$ = 0.9014. Fig. 3 shows the inversion efficiency $\delta$ versus $\Delta B_0$ and $B_1$ space for the best designs of both HSn and Tan/Tanh. The Tan/Tanh design exhibits higher inversion over a broader region of off-resonance and $B_1$ space than HSn. It also exhibits a wider transition band, which helps maintain blood pool uniformity. Therefore, we employ the Tan/Tanh design in this study.

### 3.2 Phantom studies

Table 2 summarizes the T1 values of nine vials measured with IR-FSE and coMOLLI in the phantom study, along with the corresponding accuracy and precision. The T1 estimates from coMOLLI demonstrate high accuracy, with relative errors of less than 5% compared to IR-FSE. A Bland−Altman analysis was performed to assess the agreement between coMOLLI and the reference IR-FSE method in terms of relative error. The mean relative bias was -0.72%, with 95% limits of agreement ranging from -5.89% to 4.45%. All relative errors fell within ±5% across the T1 range (350-2300 ms), indicating close agreement between coMOLLI and IR-FSE (as shown in Fig. 4).

### 3.3 In vivo studies

Myocardial T1 maps were fitted using $\delta$ = 0.855. Fig. 5 shows native myocardial T1 maps at three short-axis slices in three healthy volunteers using coMOLLI, and Supplementary Fig. S2 shows the corresponding C maps derived from the fit. The average native myocardial T1 values across 21 healthy volunteers were 1468 ± 48 ms, 1514 ± 39 ms, and 1545 ± 50 ms, and blood

T1 values were 2182 ± 132 ms, 2124 ± 153 ms, and 2131 ± 158 ms for apical, middle, and basal slices, respectively. In healthy volunteers (12 males, 9 females), global T1 values were 1523 ± 33 ms in males and 1500 ± 37 ms in females. We found no statistically significant T1 differences based on sex (P = 0.166) globally. Slice-wise mean precision across 21 healthy volunteers was 43.27 ms, 35.11 ms, and 42.71 ms for apical, middle, and basal slices, respectively.

Fig. 6 shows representative native and post Gd infusion T1w images, as well as the corresponding T1 maps acquired with coMOLLI from three patients. Post Gd infusion maps show the expected shortening of myocardial and blood T1 relative to the native values. Table 1 summarizes per-patient T1 values for the nine patients.

**4. Discussion**

This study presents the development and validation of coMOLLI, a 5T myocardial T1 mapping technique to address the challenges of conventional MOLLI. The coMOLLI method showed close agreement with IR-FSE in phantom studies and produced stable T1 maps in healthy volunteers and patients, supporting its potential clinical feasibility at 5T.

As reported in prior studies [26,29], imperfect inversion pulses can bias T1 estimation. Achieving high inversion efficiency is therefore essential for accurate myocardial T1 mapping, especially at high field. Therefore, the inversion pulse was redesigned specifically for 5T system conditions to improve the inversion efficiency. Furthermore, it was observed that optimizing the adiabatic inversion pulse not only improves the inversion efficiency but also enhances blood pool uniformity (see Supplementary Fig. S3). At 5T, shimming is typically centered on the heart, resulting in reduced field homogeneity in inflowing blood. Using a broadband Tan/Tanh inversion pulse helps maintain uniform inversion throughout local and distal blood, yielding more homogeneous T1 maps.

In addition to the redesigned inversion pulse, the inversion efficiency $\delta$ was explicitly incorporated into the fitting model. Prior methods, such as InSiL [13] and multiparametric saturation recovery single-shot acquisition (mSASHA) [30], acquire inline PDw images by adding additional readouts to the imaging sequence, enabling pixel-wise $\delta$ estimation. Although this strategy can be integrated into the proposed method, we opted to use a separate PDw scan for two main reasons. First, the additional readouts prolong acquisition and breath-hold time, which may compromise patient compliance and image quality in clinical practice. Second, estimating pixel-wise $\delta$ from inline PDw images can destabilize fits due to limited data points and increased model complexity (see Supplementary Fig. S4). Using a separate PDw scan enables more stable and efficient $\delta$ estimation, facilitating a simplified and robust three-parameter fit. We acknowledge that $\delta$ may vary across subjects and slices due to varying $B_1$. Considering the tradeoff between stability and accuracy, we choose to use a fixed $\delta$ estimated by the separate PDw scan.

Previous studies at 1.5T and 3T show mixed but recurring slice-wise patterns in healthy volunteers [31-35]. Several cohorts at 3T report higher T1 values at the apex and lower T1 values at the middle, whereas at 1.5T, the T1 differences among slices are small or not significant [32]. In our study, slice-wise native T1 differences were modest and followed a base-to-apex decreasing trend in healthy volunteers, consistent with the previous study [35]. However, given the small sample size and the greater field inhomogeneity at 5T, we consider the T1 differences among slices to be most likely attributable to field inhomogeneity, while true physiological gradients cannot be excluded. Moreover, the sex difference in global T1 at 5T was not significant, which is consistent with the previous study [35]. However, since this conclusion was drawn from a relatively small and unbalanced population, a larger and more balanced cohort with covariate adjustment is still needed to further elucidate the sex-related T1 differences at 5T.

## 5. Limitations

This study has several limitations. First, the cohort was relatively small, particularly the patient group, and a definitive pathological diagnosis was not achieved in several patients. Future work with larger and more homogeneous cohorts will be needed to further evaluate subgroup differences in myocardial T1, such as sex-related and slice-wise variations. Second, although using GRE readout is beneficial for SAR, it yields lower SNR compared to bSSFP-based T1 mapping, leading to reduced precision [36]. Denoising filters can partially alleviate this issue, but often cause image blurring. In future work, we will investigate advanced deep learning–based denoising methods to improve SNR and precision while preserving the image quality of myocardial T1 mapping at 5T. Third, we used a constant inversion efficiency to stabilize T1 model fitting in the in vivo study, disregarding its variation across subjects, slices, and tissues. In practice, however, the inversion efficiency varies with field inhomogeneity and tissue characterizations. Using a constant inversion efficiency may therefore introduce bias into the measured T1 values. For example, because blood has longer T1 and T2 values than myocardium, it exhibits a higher inversion efficiency, and using a constant value may lead to overestimation of blood T1. Conversely, in post-contrast T1 mapping, the shorter T1 corresponds to a lower inversion efficiency than in the native T1 mapping, leading to underestimation. Therefore, pixel-wise inversion efficiency fitting could further improve the accuracy of in vivo T1 quantification and will be investigated in our future work.

## 6. Conclusion

The coMOLLI method demonstrated high accuracy in phantom studies and feasibility in vivo studies. By adopting the widely used 5-(3)-3 MOLLI acquisition scheme, it shows potential for clinical cardiac imaging at 5T.

## Declarations

**Funding:** This study was supported by the National Key R&D Program of China under grant No. 2021YFF0501402 and the National Natural Science Foundation of China under grant Nos. 62322119, 62531024, U22A20344, 82020108018.

**Author contributions: Linqi Ge:** Writing – original draft, Methodology, Formal analysis, Software, Visualization, Validation, Investigation, Data curation. **Yinuo Zhao:** Writing – original draft, Methodology. **Yubo Guo:** Investigation, Validation, Data curation. **Yuanyuan Liu:** Methodology, Software. **Yihang Zhou:** Resources. **Haifeng Wang:** Resources. **Dong Liang:** Supervision, Funding acquisition. **Hairong Zheng:** Supervision, Funding acquisition. **Yining Wang:** Conceptualization, Funding acquisition, Resources. **Yanjie Zhu:** Writing – review & editing, Project administration, Conceptualization, Funding acquisition, Resources, Supervision.

**Ethics approval and consent:** The institutional ethics committee at Peking Union Medical College Hospital (Beijing, China) approved the study. All participants were required to provide written informed consent before recruitment.

**Consent for publication:** Written informed consent was obtained from all participants for inclusion of their data in publications.

**Availability of data and materials:** Anonymized datasets, which include T1 maps from healthy volunteers and patients, have been made publicly available via Zenodo (https://doi.org/10.5281/zenodo.16417553). All patient-identifiable information was removed, and the sharing protocol was approved by the institutional review board.

**Declaration of competing interests:** The authors declare that they have no known competing financial interests or personal relationships that could have appeared to influence the work reported in this paper.

**Acknowledgement:** The authors would like to acknowledge United Imaging Healthcare for their valuable support in technical assistance, which greatly contributed to this research. And the authors thank each of the study subjects for their participation.

**Tables**

**Table 1.** The specific cardiomyopathies and measured T1 values of the nine patients.

| ID | Diagnosis | Myocardium | | Blood | |
|---|---|---|---|---|---|
| | | Native T1 (ms) | Post Gd infusion T1 (ms) | Native T1 (ms) | Post Gd infusion T1 (ms) |
| P01 | Myocardial fat infiltration | 1437 | 605 | 2234 | 404 |
| P02 | Myocardial fat infiltration | 1393 | 567 | 2419 | 437 |
| P03 | Cardiac amyloidosis | 1789 | 835 | 2012 | 672 |
| P04 | NICM | 1615 | 634 | 2014 | 471 |
| P05 | NICM | 1581 | 739 | 2174 | 457 |
| P06 | NICM | 1496 | 696 | 2176 | 494 |
| P07 | NICM | 1490 | 746 | 2032 | 453 |
| P08 | MINOCA | 1509 | 722 | 2390 | 472 |
| P09 | MINOCA | 1695 | 670 | 2443 | 401 |

NICM: non-ischemic cardiomyopathy (not definitively diagnosed).

MINOCA: myocardial infarction with non obstructive coronary arteries.

**Table 2.** Phantom validation in nine vials. Reference T1 of IR-FSE and native T1, relative error, and precision of coMOLLI.

| | IR-FSE | coMOLLI | | | |
|---|---|---|---|---|---|
| No. | Reference T1 (ms) | $\delta$ | Native T1 (ms) | Relative Error (%) | Precision (ms) |
| 1 | 599.6 | 0.896 | 582.5 | 2.85 | 5.67 |
| 2 | 1501.1 | 0.915 | 1498.3 | 0.19 | 22.40 |
| 3 | 654 | 0.954 | 646.7 | 1.12 | 7.14 |
| 4 | 788.9 | 0.873 | 767.6 | 2.70 | 9.01 |
| 5 | 1872.7 | 0.908 | 1860.9 | 0.63 | 28.70 |
| 6 | 2374 | 0.971 | 2260 | 4.80 | 40.31 |
| 7 | 440.9 | 0.882 | 435.2 | 1.29 | 5.05 |
| 8 | 1230.2 | 0.892 | 1270.5 | 3.28 | 17.62 |
| 9 | 357.3 | 0.941 | 371 | 3.83 | 3.45 |

## Figure Legends

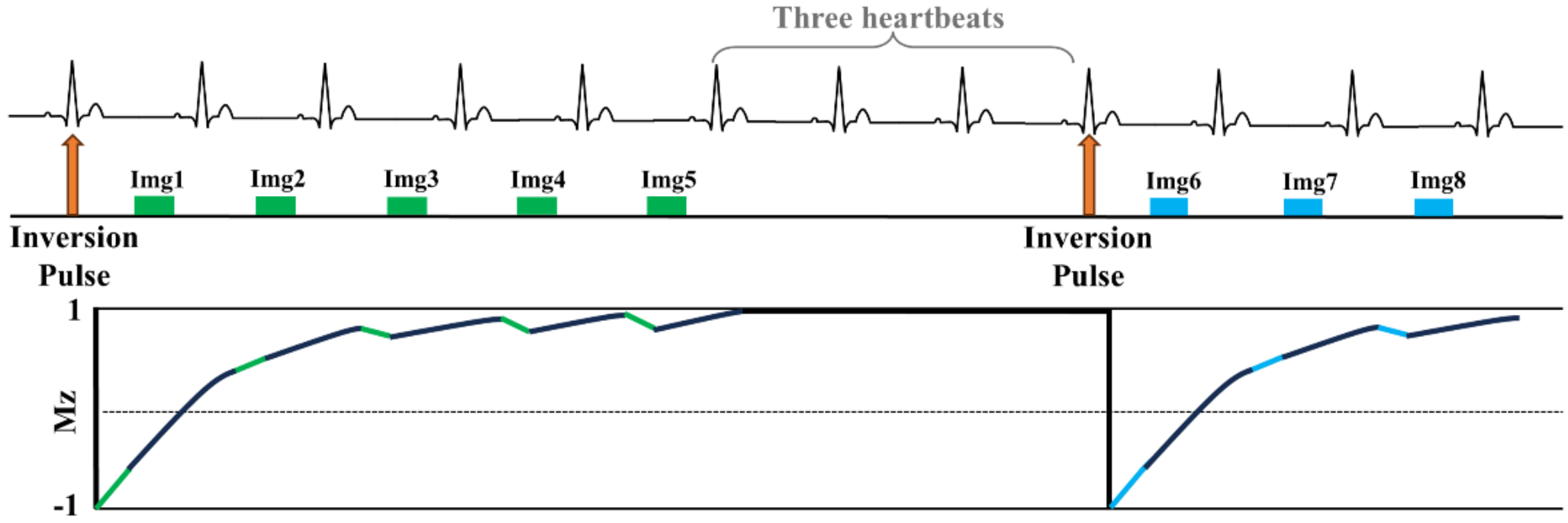

**Fig. 1.** The timing diagram of the coMOLLI sequence. The acquisition follows a 5-(3)-3 MOLLI scheme. Initially, five images (Img1−Img5) are acquired in successive mid-diastolic heartbeats after the first inversion pulse (orange arrow). A three-heartbeat rest period is then inserted to allow recovery of longitudinal magnetization before the second inversion (orange arrow), after which three additional images (Img6−Img8) are acquired. Colored bars denote GRE readouts, and the curve depicts the recovery of longitudinal magnetization ($M_z$) following each inversion pulse.

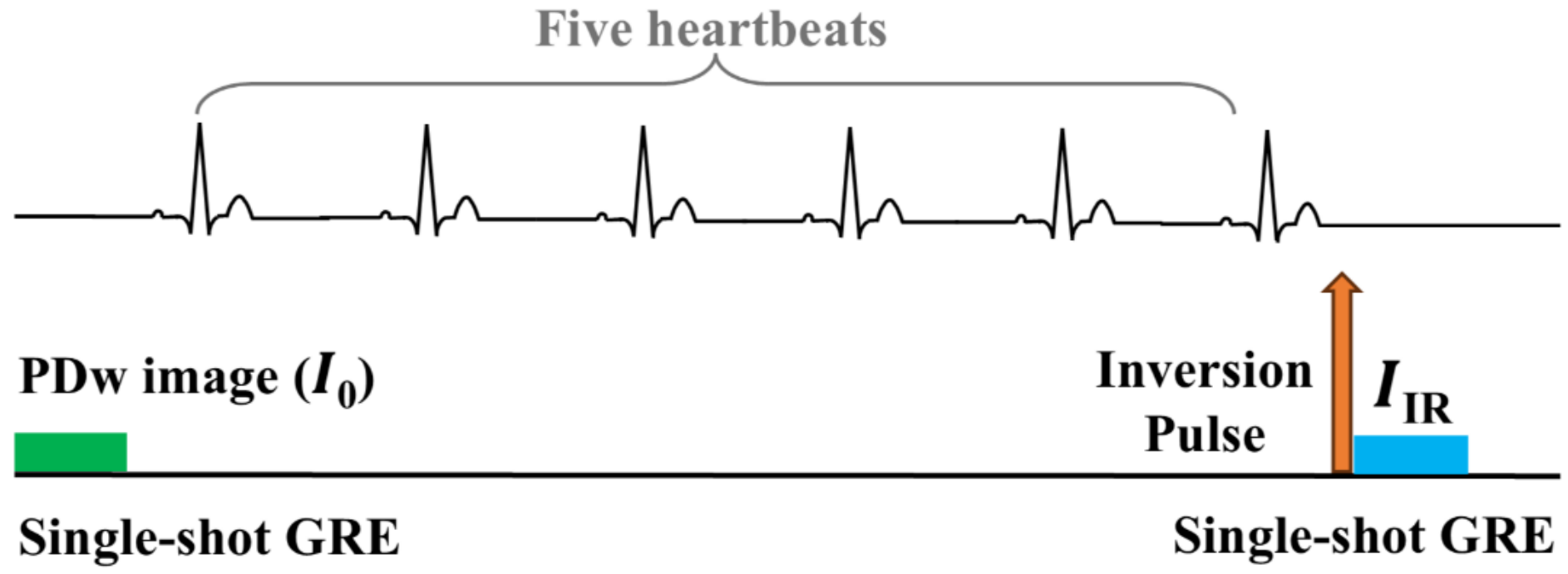

**Fig. 2.** Pre-scan for estimating the inversion efficiency $\delta$. An ECG-gated GRE scan comprising two single-shot acquisitions first obtains a PDw single-shot image ($I_0$). After a five-heartbeat rest period, an inversion pulse (orange arrow) is applied, and a second single-shot image ($I_{IR}$) is acquired immediately. Myocardial ROIs are drawn on $I_0$ and $I_{IR}$ to compute the mean signals. $\delta$ is defined as the ratio of the mean myocardial signals of $I_{IR}$ and $I_0$.

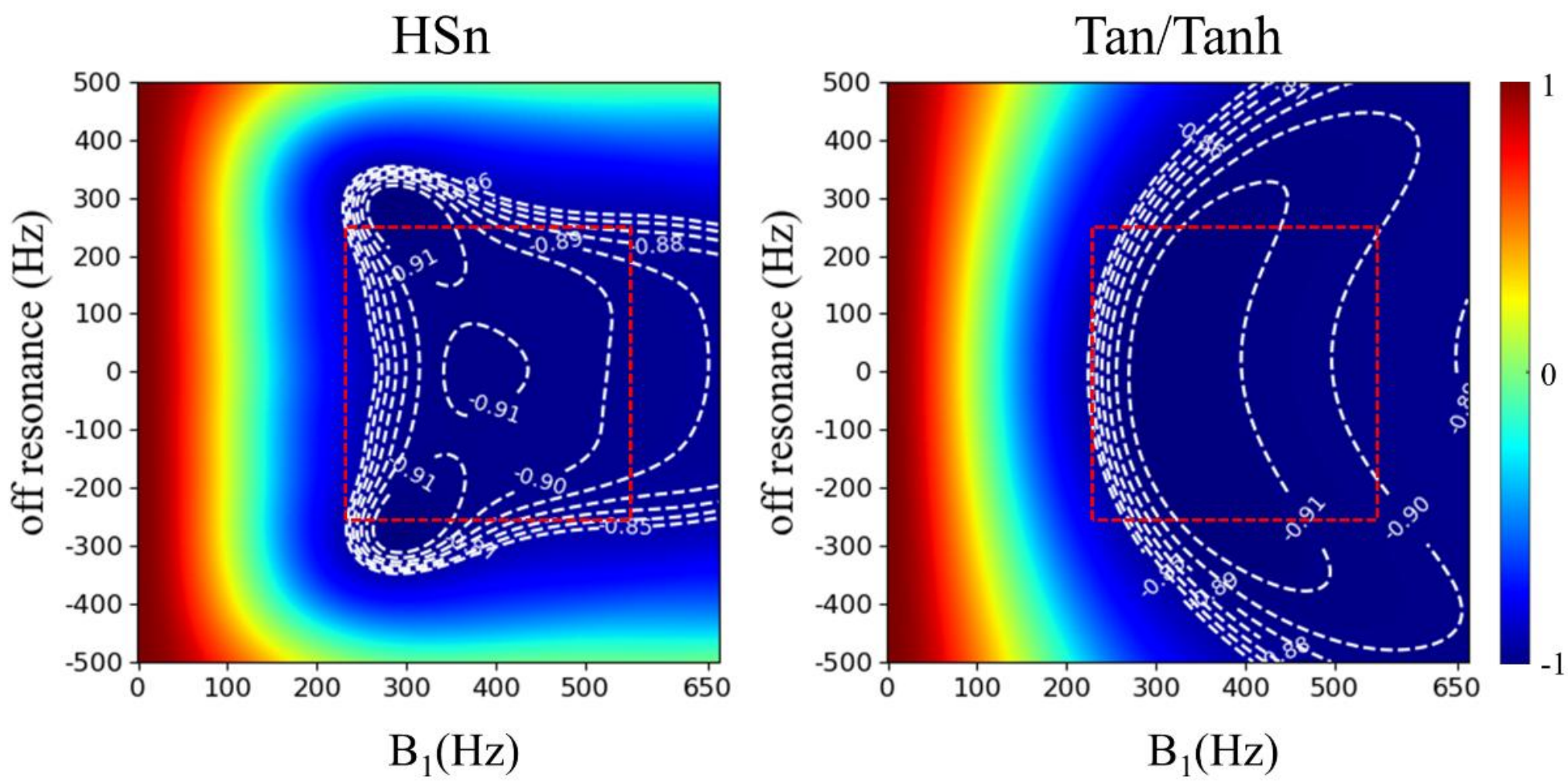

**Fig. 3.** The inversion efficiency $\delta$ maps for the optimal HSn and Tan/Tanh designs across the $B_0$ and $B_1$ ranges ($\Delta B_0$: ±250 Hz, $B_1$: 225−540 Hz). The white dashed contours denote lines of constant $\delta$, and the red dashed rectangles mark the target operating window.

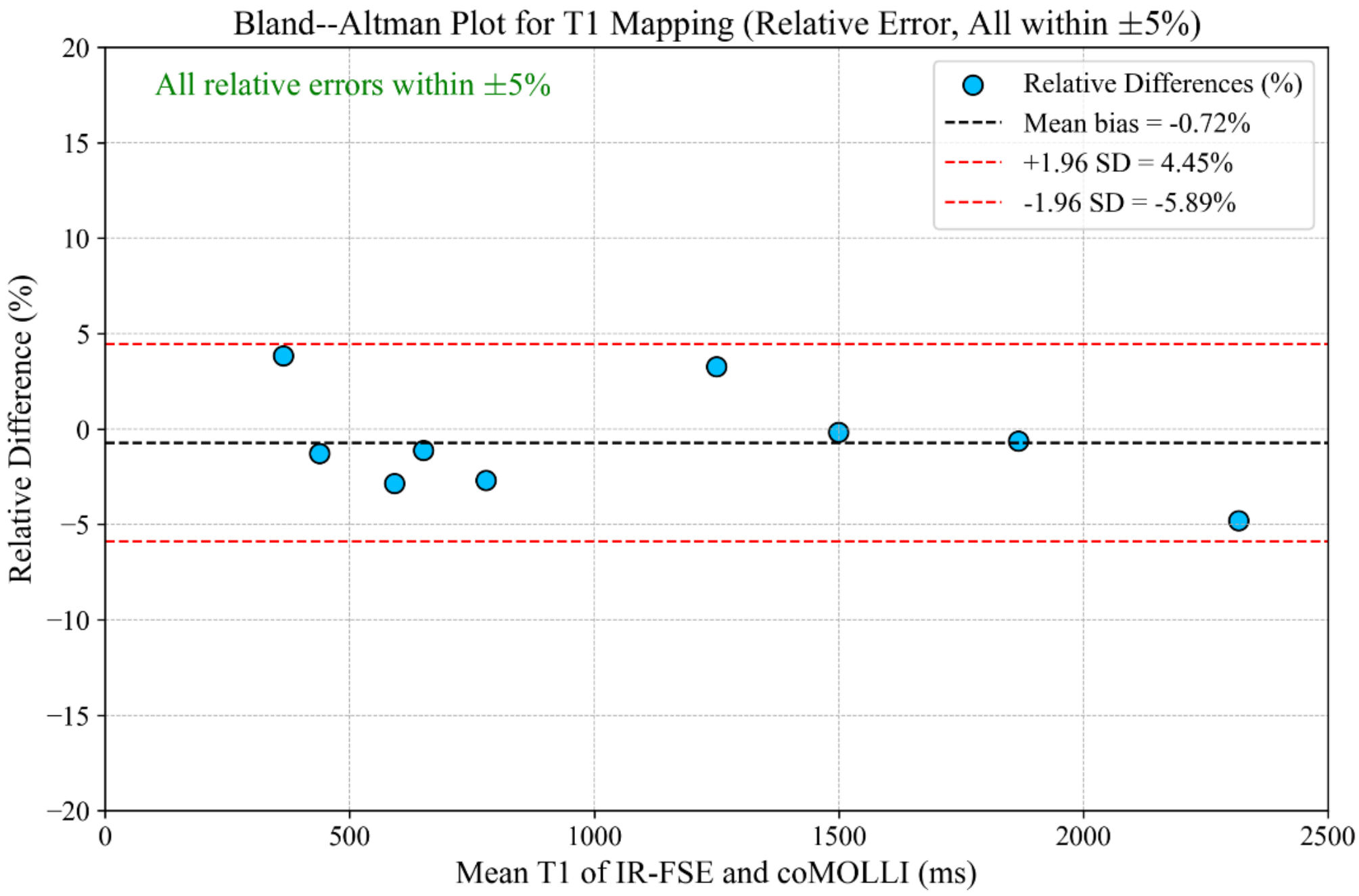

**Fig. 4.** Bland−Altman analysis of vial-wise T1 measurements in phantoms, comparing coMOLLI and IR-FSE. The blue points show the relative difference (%) plotted against the mean T1 of the two methods. The dashed black line shows the mean bias (-0.72%), and the red dashed lines show the 95% limits of agreement (-5.89% to 4.45%).

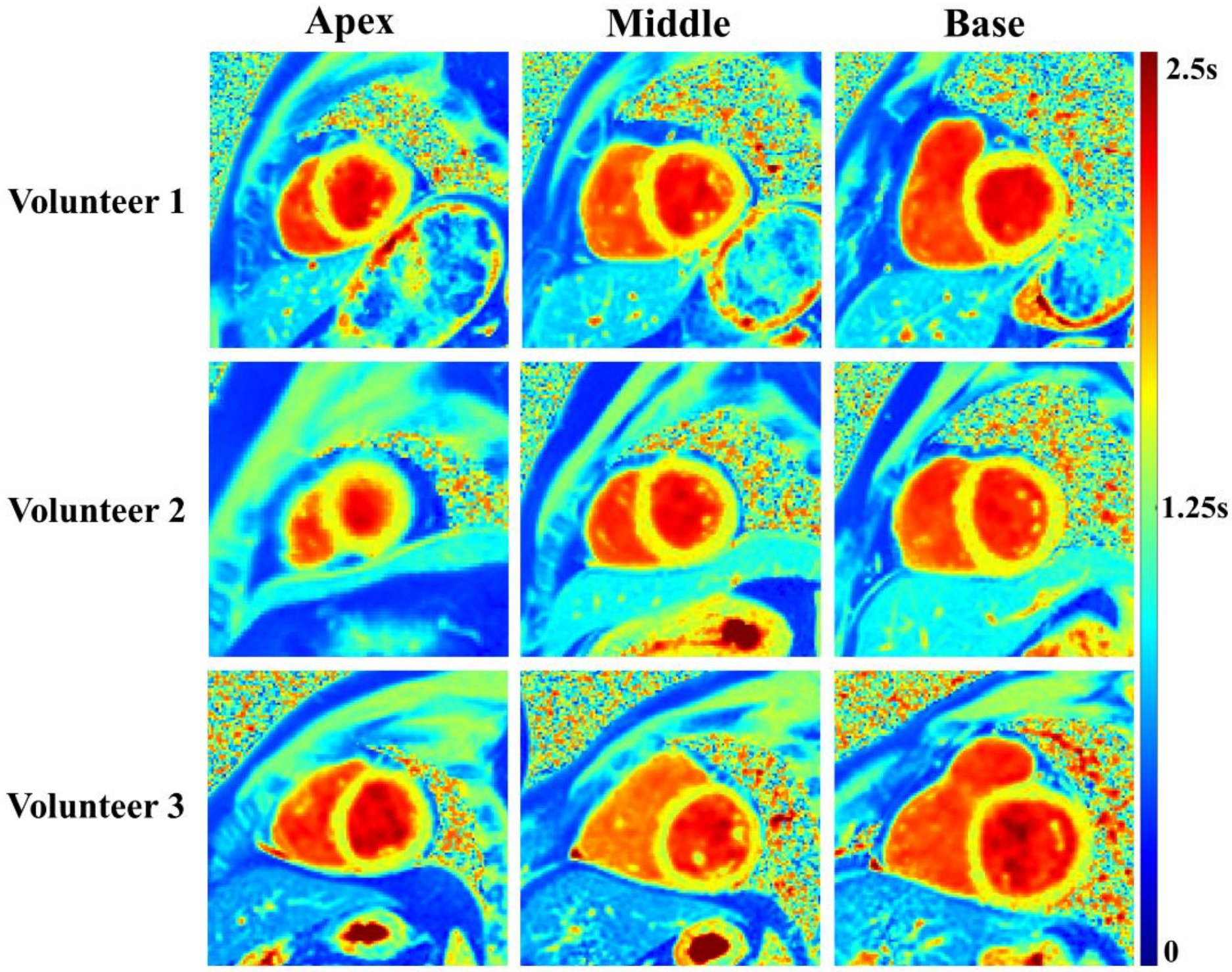

**Fig. 5.** Native myocardial T1 maps acquired with coMOLLI in three healthy volunteers at apical, middle, and basal slices. These examples illustrate the expected healthy-subject appearance, with homogeneous myocardium and a uniform blood pool signal.

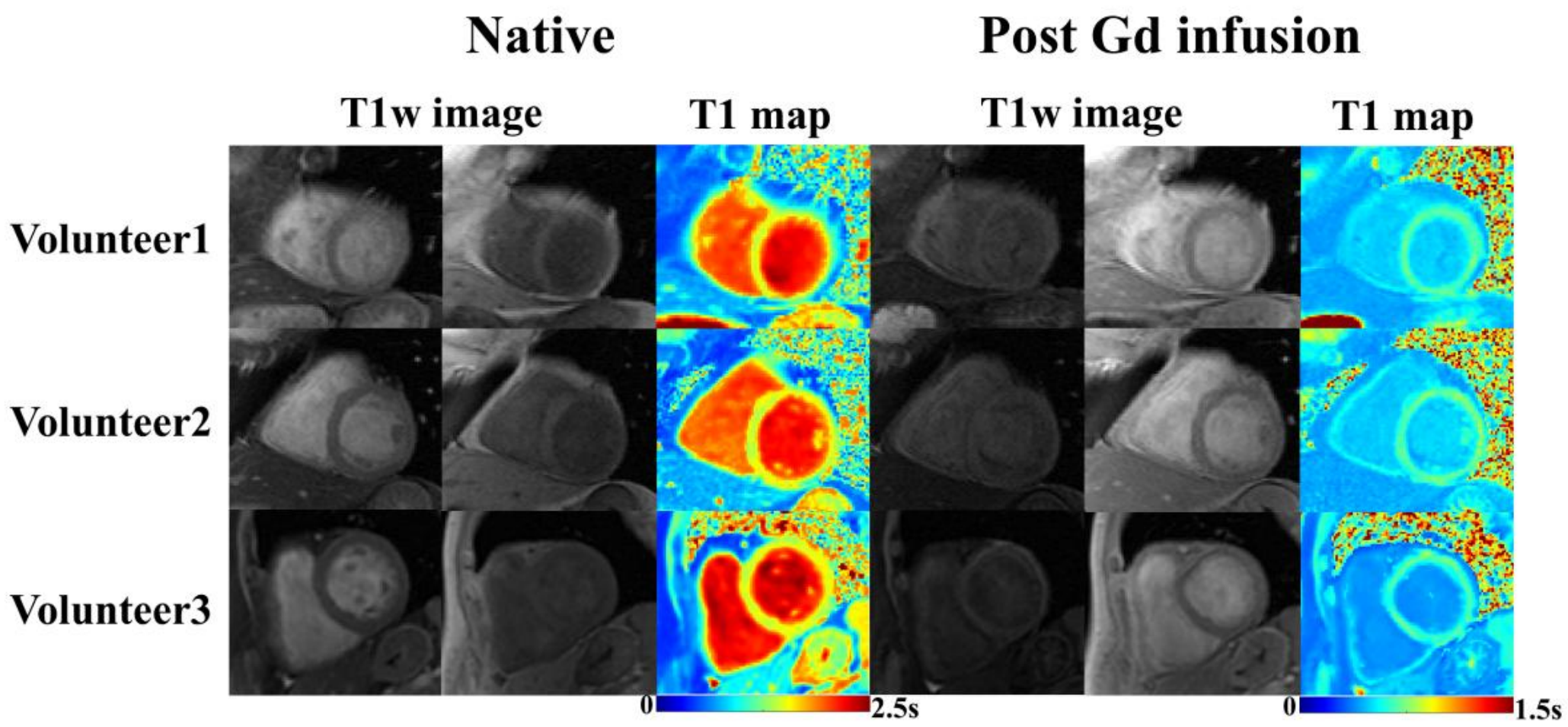

**Fig. 6.** Representative native (left) and post Gd infusion (right) T1w images and corresponding T1 maps from three patients. The T1w images are provided for anatomical reference.

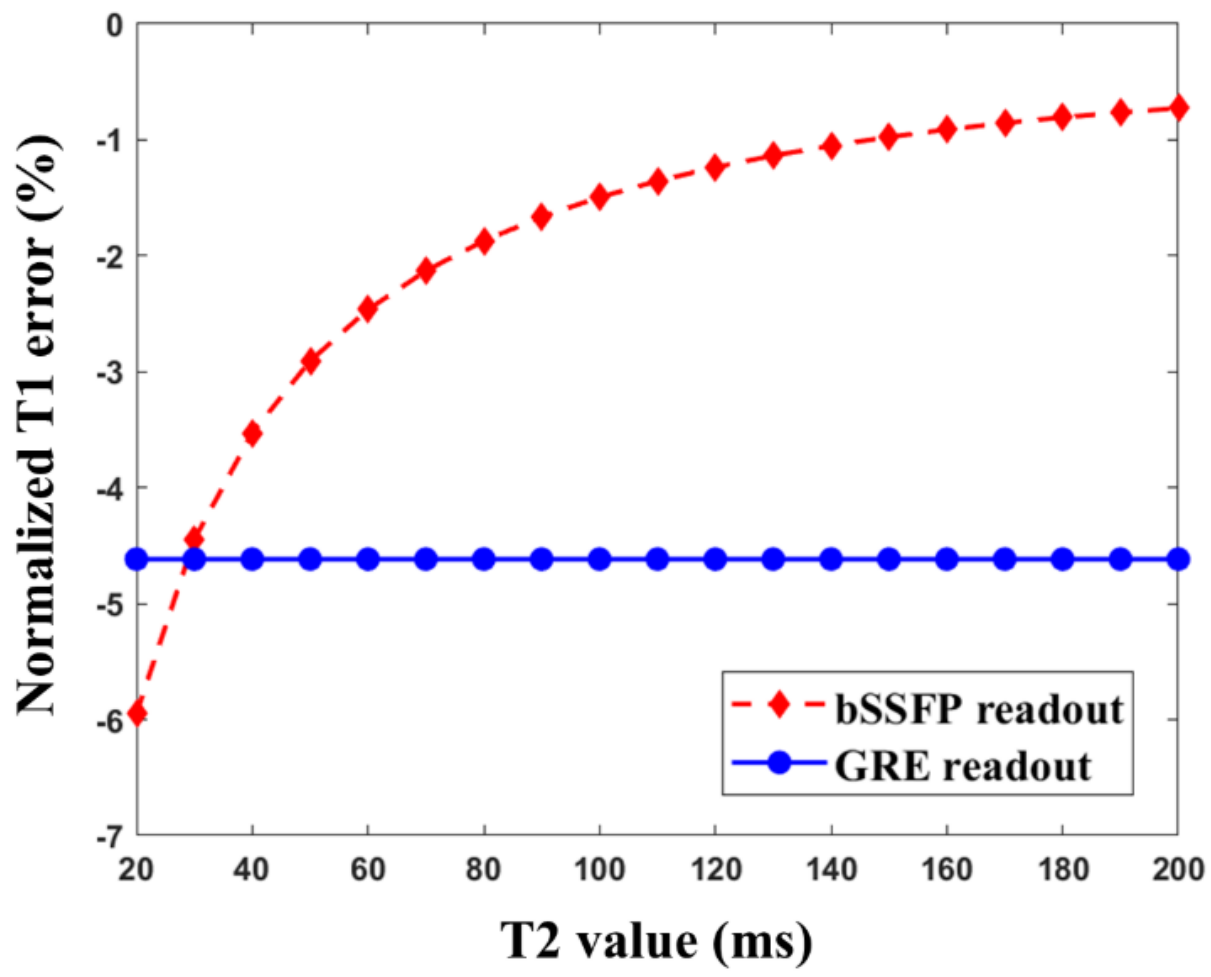

**Supplementary Fig. S1.** Normalized T1 error (%) versus T2 value (20–200 ms) from Bloch simulations. The bSSFP readout (red, dashed) exhibits a clear T2-dependent bias, while the GRE readout (blue, solid) remains essentially flat, confirming the expected T2-insensitive behavior for GRE.

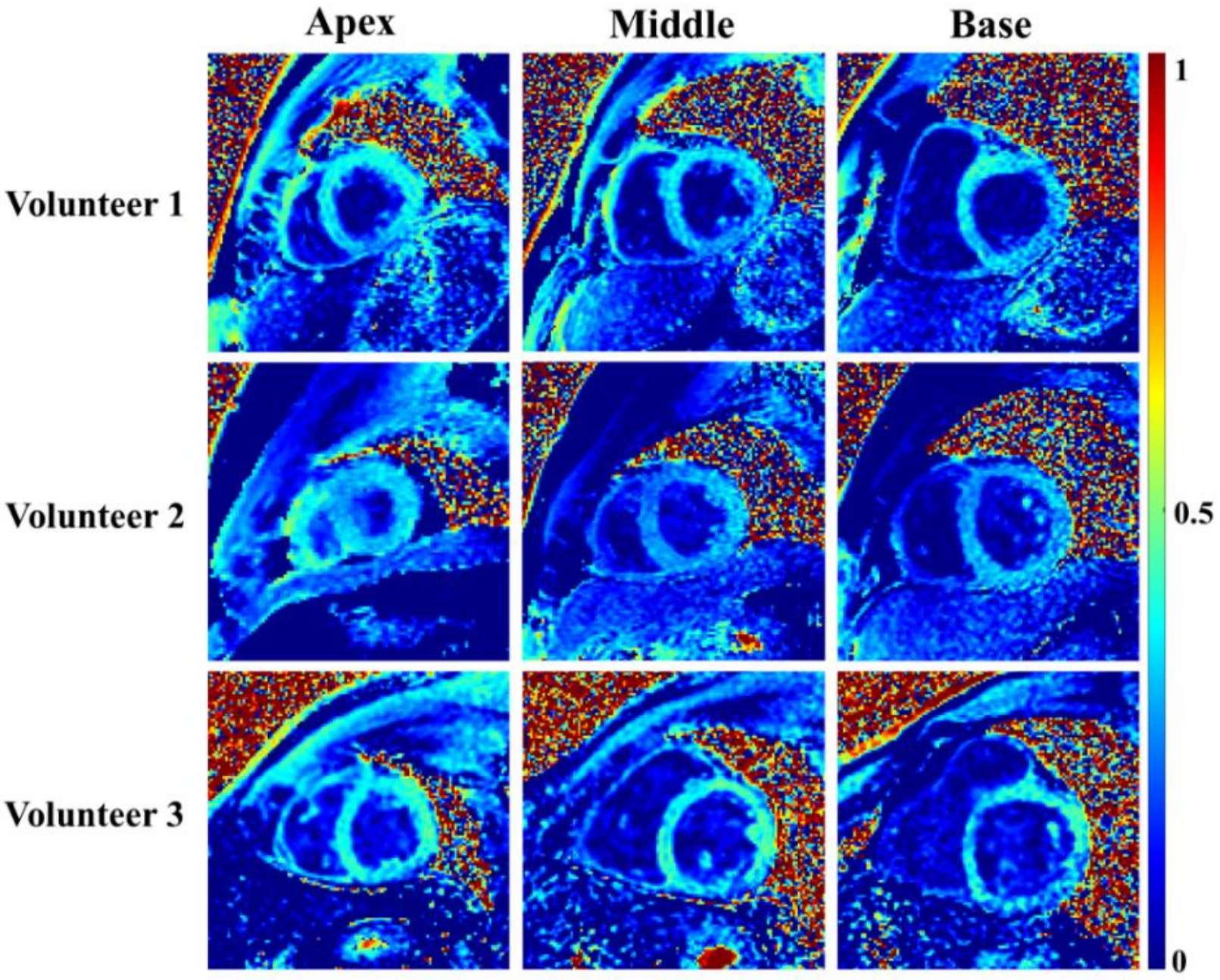

**Supplementary Fig. S2.** The $C$ maps of coMOLLI at apical, middle, and basal slices from three healthy volunteers. The global $C$ values of myocardium for the three volunteers are 0.33, 0.29, and 0.39, respectively.

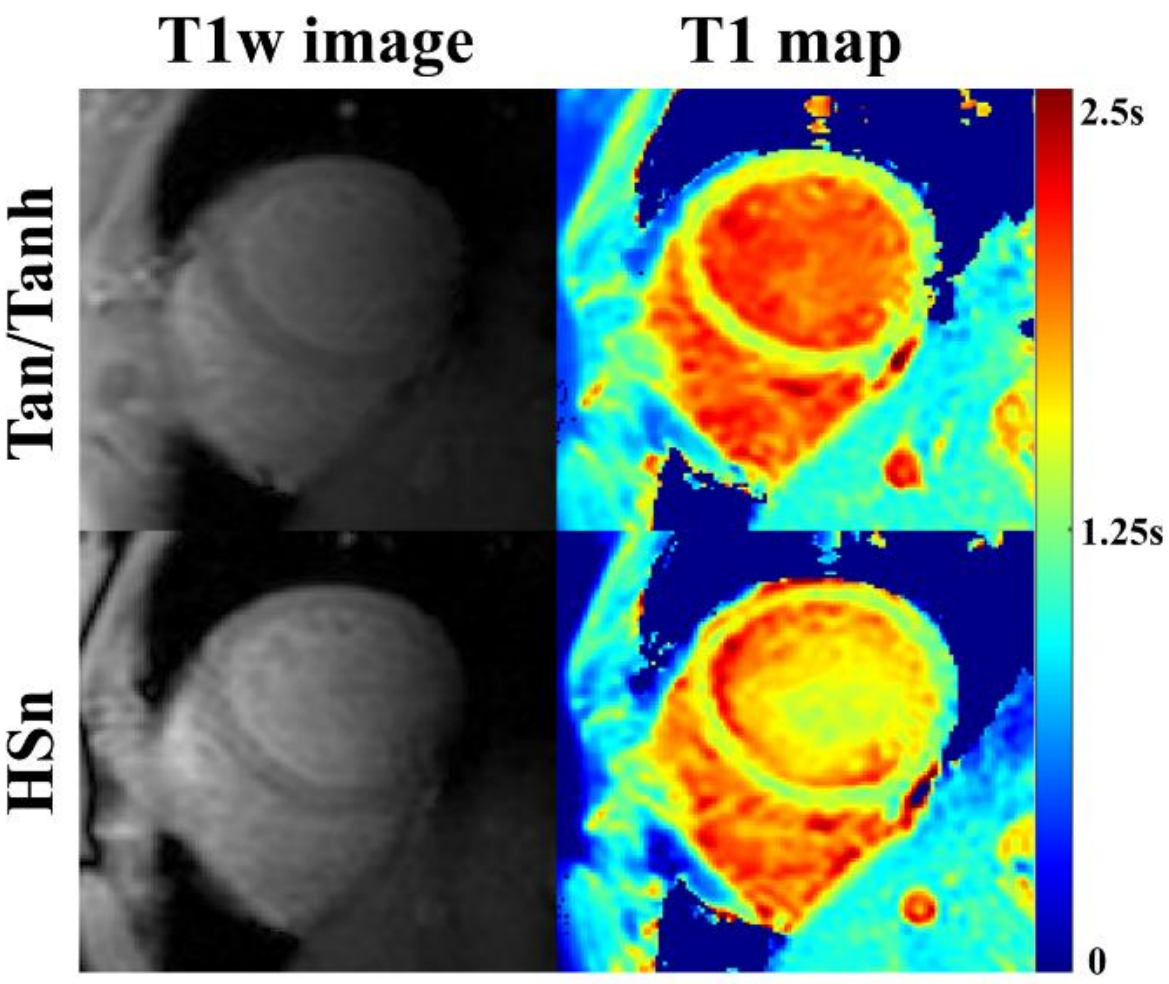

**Supplementary Fig. S3.** The corresponding T1w images and T1 maps of Tan/Tanh and HSn. From the T1 map, the blood pool uniformity of Tan/Tanh is better, while the blood flow of HSn has obvious artifacts. Moreover, by comparing the T1w image and the T1 map, the T1 value of HSn has an obvious deviation.

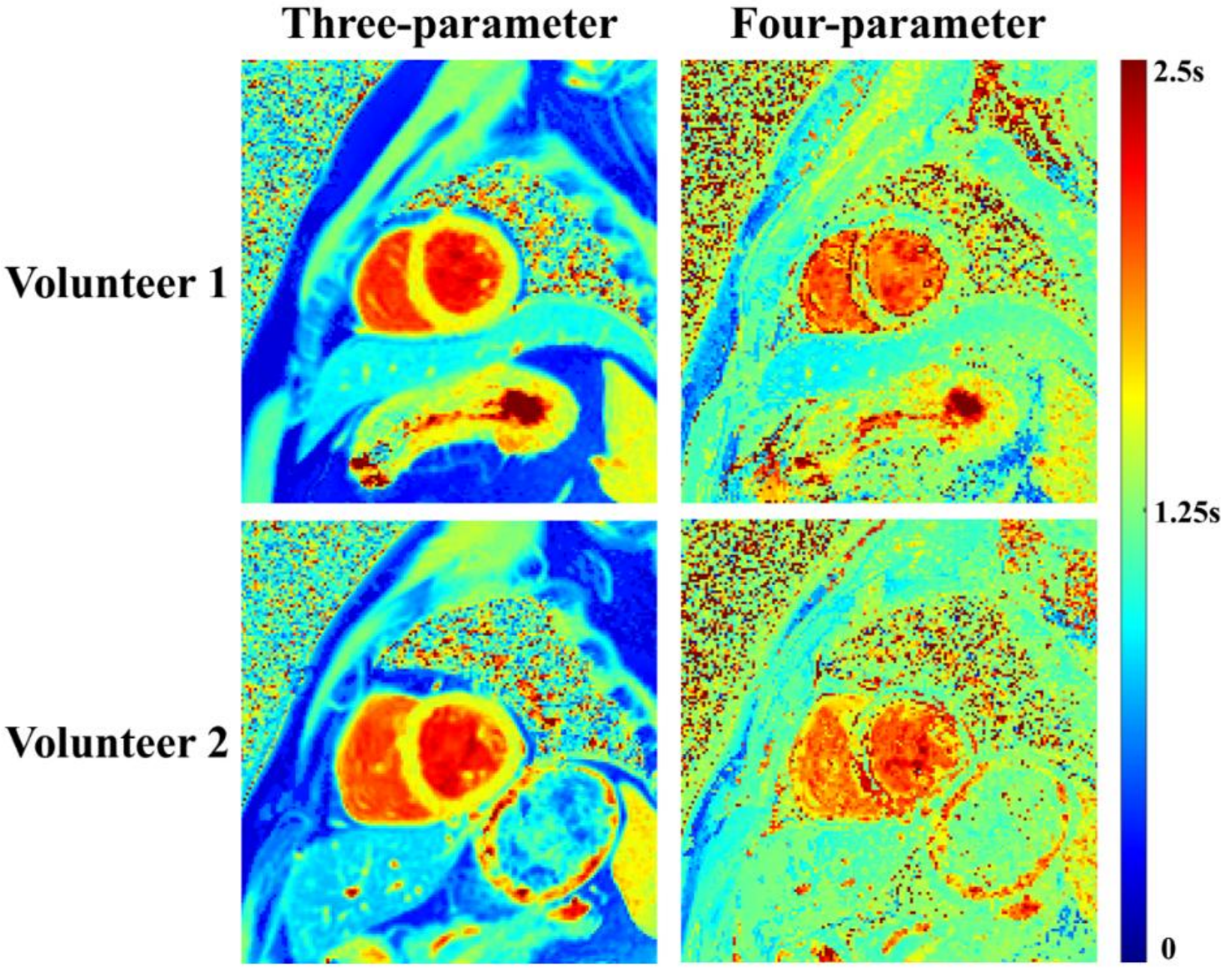

**Supplementary Fig. S4.** The T1 maps obtained by the three-parameter and four-parameter fitting models. Estimating it pixel-wise from inline PDw images can lead to unstable fitting due to limited data points and increased model complexity.